# Theoretical Evaluation of Intermolecular and Intramolecular Isotope Fractionations in Fragmentation: In the Light of Chlorine and Bromine Isotope Fractionation of Halogenated Organic Compounds Observed in Electron Ionization Mass Spectrometry

**Caiming Tang[1,2,*] and Xianzhi Peng[1,*]**

[1] *State Key Laboratory of Organic Geochemistry, Guangzhou Institute of Geochemistry, Chinese Academy of Sciences, Guangzhou 510640, China*

[2] *University of Chinese Academy of Sciences, Beijing 100049, China*

*Corresponding Author.

Tel: +86-020-85291489; Fax: +86-020-85290009; E-mail: CaimingTang@gig.ac.cn. (C. Tang).

Tel: +86-020-85290009; Fax: +86-020-85290009. E-mail: pengx@gig.ac.cn. (X. Peng).

## ABSTRACT

Revelation of the effects of isotope fractionation during fragmentation in electron ionization mass spectrometry (EI-MS) on compound-specific isotope analysis of chlorine and bromine (CSIA-Cl/Br) may be of crucial significance, yet a theoretical basis for elucidating the effects is absent. This study provides a solid theoretical deduction regarding the isotope fractionation taking place during fragmentation in EI-MS. Both intermolecular and intramolecular isotope fractionations present in dehalogenation processes, influencing the isotope ratios of detected ions. In general, intermolecular isotope fractionation positively affects the isotope ratios of precursor ions but inversely impacts those of the corresponding product ions. Molecular ions are affected by one type of normal intermolecular isotope fractionation only during the dehalogenation from parent ions to their product ions. While product ions can be impacted by two types of intermolecular isotope fractionations, namely, the inverse intermolecular isotope fractionation from parent ions to their product ions and the normal intermolecular isotope fractionation from the product ions to their further dehalogenated product ions. For a compound having position-distinct isotope ratios, if dehalogenation reacts on only one position, the intermolecular isotope fractionation taking palace in this reaction is normal for the detected precursor ion, but ineffective to the product ion isotope ratio in comparison with that of the total precursor ions. On the other hand, intramolecular isotope fractionation positively affects the isotope ratios of product ions only. The isotopologue distributions of the observed precursor ions are deduced to never obey binomial distribution, regardless of what modes of the initial isotopologue distributions of the total precursor ions are. Therefore, the binomial-theorem-based isotope ratio evaluation schemes using pair(s) of neighboring isotopologues are unlikely to obtain the isotope ratios exactly equal to those of the complete isotopologues. The isotope ratios of the complete isotopologues can be calculated with the complete-isotopologue scheme. The measured isotope ratios calculated with the isotopologue-pair scheme likely do not accurately reflect the isotope ratios in reality, because the isotopologue distribution modes of the analytes are anticipated to always be inconsistent with those of the external isotopic standards that are structurally identical to the corresponding analytes. This inference has been experimentally verified with the isotopically distinct standards of perchloroethylene and trichloroethylene from different manufacturers.

# INTRODUCTION

In the last decade, gas chromatography electron ionization mass spectrometry (GC-EI-MS) has been increasingly applied to compound-specific isotope analysis of chlorine and bromine (CSIA-Cl/Br) of halogenated organic compounds (HOCs).[1-5] GC-EI-MS can provide comparably precise and accurate CSIA-Cl results as the conventional offline isotope ratio MS (IRMS) [Aeppli et al, C 2010], but is much more simple, efficient and cost-effective [Bernstein et al, 2011]. GC online IRMS (GC-IRMS) can also be used for CSIA-Cl [Shouakar-Stash et al, 2006], but only provide high precision for a narrow range of organochlorines; while GC-EI-MS can be applied to more universal compounds [Bernstein et al, 2011]. Due to the obvious advantages and promising prospective, CSIA-Cl/Br using GC-EI-MS has raised increasing concerns recently [Palau et al, 2014; Liu et al, 2016; Heckel et al, 2017].

Offline IRMS converts all chlorine atoms of an analyte into singly chlorinated molecules such as $CH_3Cl$ and CsCl for CSIA-Cl [Jendrzejewski et al, 1997; Westaway et al, 1998; Holmstrand et al, 2004; Holt et al, 2001; Shouakar-Stash et al, 2006; Holt et al, 1997; Numata et al, 2002]. While CSIA-Cl using GC-EI-MS measures isotope ratios based on the calculation schemes using molecular ions and/or fragmental ions [Jin et al, 2011; Elsner et al, 2008]. In GC-EI-MS, fragmentation occurs inevitably during ionization and metastable-ion dissociation processes. It has been reported that “staggering large” hydrogen/deuterium isotope effects could present during fragmentation in EI-MS [Derrick, 1983], with the kinetic isotope effects (KIEs) reaching up to $10^2$-$10^3$ [Ottinger et al, 1965; Löhle et al, 1969]. Both intermolecular and intramolecular isotope effects could present in fragmentation, and the intramolecular KIEs can be approximately evaluated with the relative abundances of fragmental ions [Derrick, 1983]. In a recent study, we observed Cl/Br isotope fractionation of a number of HOCs occurring during fragmentation in EI-MS [Tang et al, 2017]. The in-fragmentation isotope fractionation in EI-MS could affect the CSIA results [Sakaguchi-Söder et al, 2007; Aeppli et al, 2010], thus the external isotopic standards with the identical structures as the analytes are necessary [Bernstein et al, 2011; Palau, & Cretnik et al, 2014; Heckel et al, 2017].

UP to now, it is still unclear how the in-fragmentation isotope fractionation affect the relative abundances of the isotopologue ions which are employed to calculate isotope ratios, neither how it further impact the directly measured isotope ratios by EI-MS prior to calibration using external isotopic standards. The main objectives of this study are to ascertain these issues, and to reveal the reasons why the external isotopic standards that are structurally identical to the analytes are necessary for CSIA using EI-MS. We also investigate that the presently generally used isotope ratio evaluation schemes using isotopologue pair(s) of fragmental ions and/or molecular ions to what extents are reasonable. Implications are proposed in order to obtain precise and accurate results for CSIA using EI-MS, e.g., using the complete-isotopologue calculation scheme of isotope ratio with whole molecular-ion isotopologues, and application of the ion source capable of generating sufficiently stable EI energies.

## PART 1: INTERMOLECULAR ISOTOPE FRACTIONATION

Actually, the isotope fractionation occurring in EI-MS is a behavior of ions rather than molecules. However, the inferences and conclusions obtained in this study will be not only applicable to ions, but also to other species and other forms of compounds. Therefore, we use the terms "intermolecular" and "intramolecular", instead of "inter-ion" and "intra-ion", respectively.

We hypothesize that the fragmentation processes of HOCs in EI-MS follows the following sequence: molecular ion → losing Cl/Br atom(s) one by one → a series of product ions containing successive numbers of Cl/Br atom(s). We define an ion with *n* Cl/Br atoms as the precursor ion of the ion with *n*-1 Cl/Br atom(s), and their carbon skeletons remain intact; and the latter was defined as the product ion of the former. In addition, we define the precursor ion which is finally transformed into the product ion as the parent ion of the product ion.

In this section, we do not take the intramolecular isotope fractionation into account for more ambitious elucidation of intermolecular isotope fractionation.

We conceive a compound in possession of four position-equivalent chlorine atoms ($Cl_4$) and take it for an example to elucidate intermolecular chorine isotope fractionation (Figure 1). Moreover, after the loss of one Cl atom, we hypothesize that the remaining three Cl atoms on the dechlorinated radical fragment of the compound are still position-equivalent. This compound has five theoretical isotopologues. For simplifying the isotopologue formulas, we omitted the carbon and other unconcerned elements, such as hydrogen and oxygen. Thus, the five isotopologue formulas can be written as $^{35}Cl_4$, $^{35}Cl_3{}^{37}Cl$, $^{35}Cl_2{}^{37}Cl_2$, $^{35}Cl^{37}Cl_3$, and $^{37}Cl_4$, as illustrated in Figure 1. If intramolecular isotope fractionation is not taken into consideration, then the generation reactions of the product ions from the molecular ions are nonselective (not isotopically selective and position-selective), and the corresponding probabilities can be theoretically calculated as shown in Figure 1.

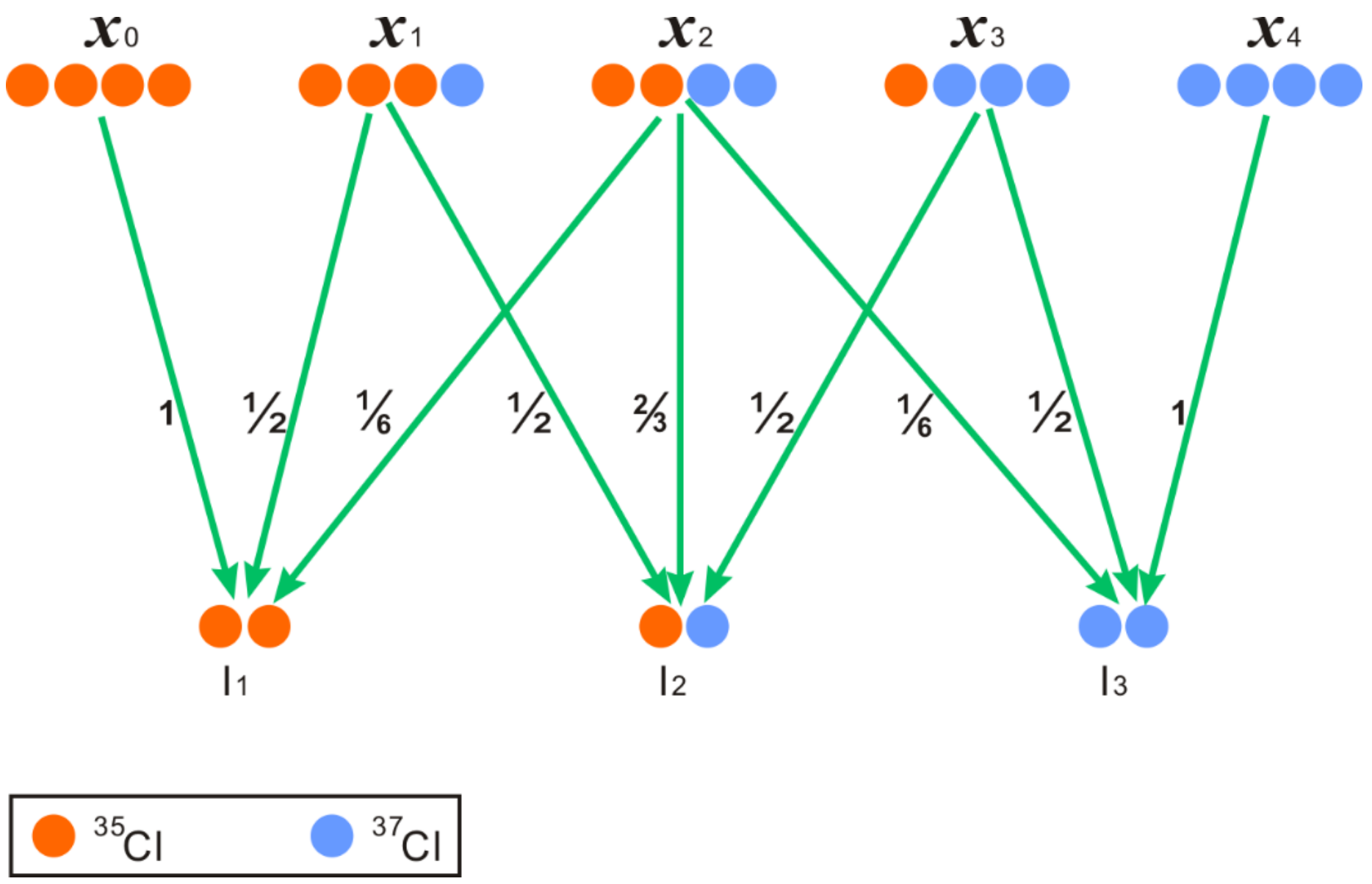

**Figure 1.** Illustration of pathways and the probabilities from molecular ion isotopologues to dechlorinatied product ion isotopologues in EI-MS. $x_0$-$x_4$: molar amounts of the molecular ion isotopologues; $I_1$-$I_3$: MS signal intensities of the product ion isotopologues.

For a molecular ion with $n$ Cl atoms ($Cl_n$), if an isotopologue has $i$ $^{37}Cl$ atoms, then the probability (P) of a product ion isotopologue which is derived from the molecular ion by losing $r$ Cl atoms and in possession of $t$ $^{37}Cl$ atoms is:

$$P(n,i,t,r) = \frac{C_{n-i}^{n-r-t} C_i^t}{C_n^r} \qquad (1.1)$$

For the compound $Cl_4$, the probability of the product ion isotopologue produced from the molecular ion by losing two Cl atoms, and possessing $t$ $^{37}Cl$ atoms is:

$$P(4,i,t,2) = \frac{C_{n-i}^{n-2-t} C_i^t}{C_4^2} \qquad (1.2)$$

Then the probabilities of the product ion isotopologues from the molecular ion isotopologues can be obtained (Figure 1). We hypothesize the molar amounts of the molecular ion isotopologues as $x_0$, $x_1$, $x_2$, $x_3$ and $x_4$, and the MS signal intensities of the product ion isotopologues as $I_1$, $I_2$ and $I_3$; if the proportionality constant of the molar amounts of the ions relative to their MS signal intensities was *k*, then the following equations can be obtained:

$$1\cdot x_0 + \tfrac{1}{2}\cdot x_1 + \tfrac{1}{6}\cdot x_2 + 0\cdot x_3 + 0\cdot x_4 = \tfrac{1}{k} I_1 \qquad (1.3)$$

$$0\cdot x_0 + \tfrac{1}{2}\cdot x_1 + \tfrac{2}{3}\cdot x_2 + \tfrac{1}{2}\cdot x_3 + 0\cdot x_4 = \tfrac{1}{k} I_2 \qquad (1.4)$$

$$0\cdot x_0 + 0\cdot x_1 + \tfrac{1}{6}\cdot x_2 + \tfrac{1}{2}\cdot x_3 + 1\cdot x_4 = \tfrac{1}{k} I_3 \qquad (1.5)$$

Let eq 1.3 × 2 + eq 1.4, we have

$$2\cdot x_0 + \tfrac{3}{2}\cdot x_1 + 1\cdot x_2 + \tfrac{1}{2}\cdot x_3 + 0\cdot x_4 = \tfrac{2}{k} I_1 + \tfrac{1}{k} I_2 \qquad (1.6)$$

And let eq 1.5 × 2 + eq 1.4, we have

$$0\cdot x_0 + \tfrac{1}{2}\cdot x_1 + 1\cdot x_2 + \tfrac{3}{2}\cdot x_3 + 2\cdot x_4 = \tfrac{2}{k} I_3 + \tfrac{1}{k} I_2 \qquad (1.7)$$

Let eq 7 / eq 6, we have

$$\frac{0\cdot x_0 + 1\cdot x_1 + 2\cdot x_2 + 3\cdot x_3 + 4\cdot x_4}{4\cdot x_0 + 3\cdot x_1 + 2\cdot x_2 + 1\cdot x_3 + 0\cdot x_4} = \frac{2I_3 + I_2}{2I_1 + I_2} \qquad (1.8)$$

Accordingly, we conceived a general equation by expanding eq 1.8 to

$$\frac{\sum_{i=0}^{n} i x_i}{\sum_{i=0}^{n} (n-i) x_i} = \frac{\sum_{i=0}^{n}\sum_{t=0}^{n-r} t C_{n-i}^{n-r-t} C_i^t x_i}{\sum_{i=0}^{n}\sum_{t=0}^{n-r} (n-r-t) C_{n-i}^{n-r-t} C_i^t x_i} \qquad (1.9)$$

where $n$ is the number of Cl/Br atoms of a molecular ion; $i$ is the number of $^{37}$Cl or $^{81}$Br atom(s) of an isotopologue of the molecular ion; $r$ is the number of the lost Cl/Br atom(s); $t$ is the number of $^{37}$Cl or $^{81}$Br atom(s) of an isotopologue of a product ion derived from molecular ion; $x_i$ is the molar amount of isotopologue $i$ of the molecular ion.

Eq 1.9 can be mathematically proved as follows.

We define

$$IR_{par} = \frac{\sum_{i=0}^{n} i x_i}{\sum_{i=0}^{n} (n-i) x_i} \quad (1.10)$$

$$IR_{pro} = \frac{\sum_{i=0}^{n}\sum_{t=0}^{n-r} t C_{n-i}^{n-r-t} C_i^t x_i}{\sum_{i=0}^{n}\sum_{t=0}^{n-r} (n-r-t) C_{n-i}^{n-r-t} C_i^t x_i} \quad (1.11)$$

where $IR_{par}$ and $IR_{pro}$ are the isotope ratios of a parent ion and the corresponding product ion, respectively.

Proof:

$$IR_{par} = IR_{pro}$$

**Proof process:**

Functions $F(i, n)$ and $G(i, n)$ as were defined as:

$$F(i,n) = \sum_{t=0}^{n-r} t C_{n-i}^{n-r-t} C_i^t \quad (1.12)$$

$$G(i,n)=\sum_{t=0}^{n-r}(n-r-t)C_{n-i}^{n-r-t}C_i^t \qquad (1.13)$$

*F* (*i*, *n*) simplifies to

$$F(i,n,r)=\sum_{t=0}^{n-r} tC_{n-i}^{n-r-t}C_i^t(i-r\le t\le i,0\le t\le n-r)$$
$$=\sum_{t=0}^{r}(i-r+t)C_{n-i}^{t}C_i^{r-t}$$
$$=\sum_{t=0}^{r}(i-r+t)C_{n-i}^{t}\frac{i!}{(r-t)!(i-r+t)!} \qquad (1.14)$$
$$=i\sum_{t=0}^{r}C_{n-i}^{t}\frac{(i-1)!}{(r-t)!(i-1-(r-t))!}$$
$$=i\sum_{t=0}^{r}C_{n-i}^{t}C_{i-1}^{r-t}$$

We can obtain the following equation according to combination principles:

$$\sum_{t=0}^{r}C_{n-i}^{t}C_{i-1}^{r-t}=C_{n-1}^{r} \qquad (1.15)$$

And *G* (*i*, *n*) simplifies to

$$G(i,n)=\sum_{t=0}^{n-r}(n-r-t)C_{n-i}^{n-r-t}C_i^t(i-r\le t\le i,0\le t\le n-r)$$
$$=\sum_{t=0}^{r}(n-i-t)C_{n-i}^{t}C_i^{r-t}$$
$$=\sum_{t=0}^{r}(n-i-t)\frac{(n-i)!}{t!(n-i-t)!}C_i^{r-t} \qquad (1.16)$$
$$=(n-i)\sum_{t=0}^{r}\frac{(n-i-1)!}{t!(n-i-1-t)!}C_i^{r-t}$$
$$=(n-i)\sum_{t=0}^{r}C_{n-i-1}^{t}C_i^{r-t}$$

Similarly, we can obtain the following equation:

$$\sum_{t=0}^{r} C_{n-i-1}^{t} C_{i}^{r-t} = C_{n-1}^{r} \quad (1.17)$$

Substituting eqs 1.14, 1.15, 16 and 1.17 into eq 1.10, we have

$$IR_{pro} = \frac{\sum_{i=0}^{n}\sum_{t=0}^{n-r} t C_{n-i}^{n-r-t} C_{i}^{t} x_i}{\sum_{i=0}^{n}\sum_{t=0}^{n-r} (n-r-t) C_{n-i}^{n-r-t} C_{i}^{t} x_i}$$

$$= \frac{\sum_{i=0}^{n} i C_{n-1}^{r} x_i}{\sum_{i=0}^{n} (n-i) C_{n-1}^{r} x_i} \quad (1.18)$$

$$= \frac{\sum_{i=0}^{n} i x_i}{\sum_{i=0}^{n} (n-i) x_i}$$

$$= IR_{par}$$

Thus, $IR_{par} = IR_{pro}$ is proved.

Therefore, for the dehalogenation process from a parent ion to the product ion, the number ($r$) of the lost Cl/Br atom is 1. Then, according to eq 1.9, we have

$$\frac{\sum_{i=0}^{n} i x_i}{\sum_{i=0}^{n} (n-i) x_i} = \frac{\sum_{i=0}^{n}\sum_{t=0}^{n-1} t C_{n-i}^{n-1-t} C_{i}^{t} x_i}{\sum_{i=0}^{n}\sum_{t=0}^{n-1} (n-1-t) C_{n-i}^{n-1-t} C_{i}^{t} x_i} \quad (1.19)$$

where $n$ is the number of Cl/Br atoms of a parent ion; $i$ is the number of $^{37}$Cl or $^{81}$Br atom(s) of an isotopologue of the parent ion; $t$ is the number of $^{37}$Cl or $^{81}$Br atom(s) of an isotopologue of the product ion; $x_i$ is the molar amount of isotopologue $i$ of the parent ion. It is notable that this equation is applicable to the parent ion whose isotope atoms are all position-equivalent, but not suitable to those with position-distinct isotope atoms.

As shown in eq 1.19, the left algebraic fraction is the isotope ratio of the parent ion, and the right is that of the product ion. This equation demonstrates that the isotope ratio of the product ion is identically equal to that of the parent ion, if intramolecular isotope fractionation does not take place or is not taken into consideration. As the light isotopologues of a precursor ion are more liable to lose an isotope atom compared with the heavy ones, the heavy isotope thus accumulates in the remaining isotopologues of the precursor ion, and is diluted in the isotopologues of the product ion. Therefore, the intermolecular isotope fractionation can normally affect the isotope ratio of the precursor ion (enrichment in heavy isotope), but inversely impact the isotope ratio of the product ion (enrichment in light isotope). Accordingly, the measured isotope ratio of a molecular ion (unfragmented molecular ions) is supposed to be higher than that of all the ionized molecular ions (both fragmented and non-fragmented), of which the isotope ratio is deduced to exceed that of fragmented molecular ion (parent ion). The isotope ratio of a product ion is impacted by two types of intermolecular isotope fractionations from the parent ion to the product ion and during the further dehalogenation from this product ion to its next-step product ion. The former influences the measured isotope ratio of this product ion inversely, while the latter affect the measured isotope ratio normally.

For an asymmetric compound possessing position-distinct isotope atoms, if the dehalogenation process from a parent ion to the product ion occurs on only one position, then the remaining isotope atoms on the dehalogenated fragment have the isotope ratio equivalent to that of the total precursor ions (both fragmented and non-fragmented), provided that the isotope atoms on the reacting position have the isotope ratio identical to that on the non-reacting positions. The isotope ratio of the product ion is deduced to be less than that of the remaining (non-fragmented) precursor ion, but exceed that of the parent ion. Therefore, the intermolecular isotope fractionation occurring in this condition is normal for the precursor ion, but with no effect to the product ion in comparison with the total precursor ions.

A prerequisite for the intermolecular to happen is that the molar amount of precursor ion should exceed that of the parent ion. In other words, the precursor ion should not be completely dehalogenated to give rise to the product ion during one-step dehalogenation process. Otherwise, the isotope ratio of the product ion will be equal to that of the precursor ion according to eq 19. In this situation, the precursor ion cannot be observed in EI-MS.

## PART 2: INTRAMOLECULAR ISOTOPE FRACTIONATION

In this part, we only focus on the dehalogenation process from a parent ion to the corresponding product ion.

We conceived a fictitious molecule (M0) containing 2 Cl atoms as illustrated in Figure 2. The symmetry (structural equivalence) of bonds a and b of M0 is indeterminate. M0 has four possible molecular isotopomers entitled as MI1, MI2, MI3, and MI4, whose structures are depicted in Figure 2. The molar amount of M0 is $a_1+a_2+a_3+a_4$, and those of MI1, MI2, MI3, and MI4 were $a_1$, $a_2$, $a_3$ and $a_4$, respectively. The chlorine isotope-related bond of the four isotopomers are marked as $a_1$', $b_1$', $a_2$', $b_2$', $a_3$', $b_3$', $a_4$', and $b_4$', respectively.

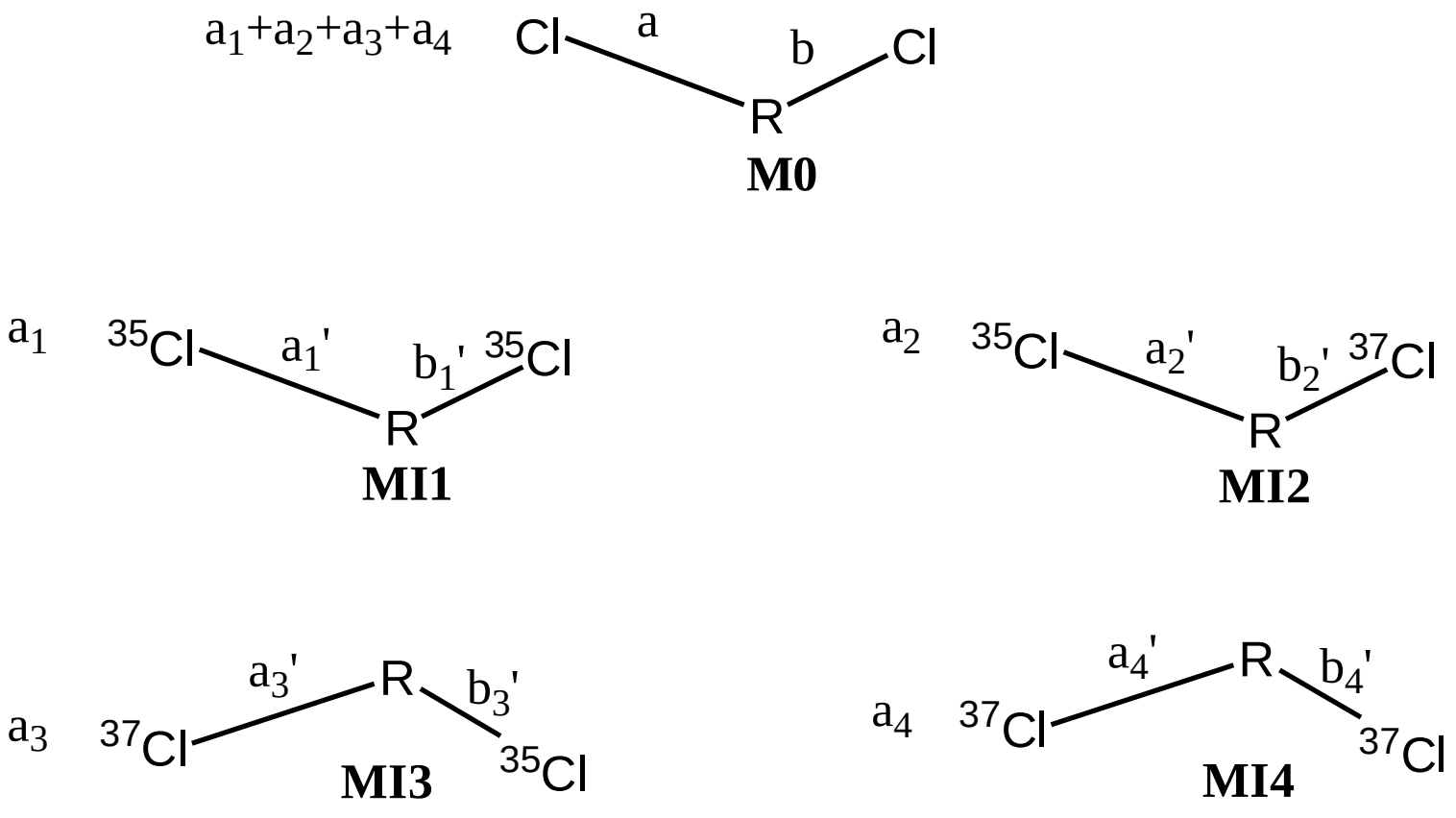

**Figure 2**. The structures of the fictitious molecule (M0) containing two chlorine atoms and the four possible isotopomers (MI1, MI2, MI3 and MI4). a: bond a of M0, b: bond b of M0; $a_1$': bond $a_1$' of M1, $b_1$': bond $b_1$' of M1; $a_2$': bond $a_2$' of M2, $b_2$': bond $b_2$' of M2; $a_3$': bond $a_3$' of M3, $b_3$': bond $b_3$' of M3; $a_4$': bond $a_4$' of M4, $b_4$': bond $b_4$' of M4. The molar amounts of M0, MI1, MI2, MI3 and MI4 are $a_1+a_2+a_3+a_4$, $a_1$, $a_2$, $a_3$ and $a_4$, respectively.

M0 has three theoretical isotopologues, and the isotopomers MI2 and MI3 share the same isotopologue ($^{35}Cl^{37}Cl$, unconcerned elements omitted). The isotopologues of MI1 and MI4 are different ($^{35}Cl_2$ and $^{37}Cl_2$).

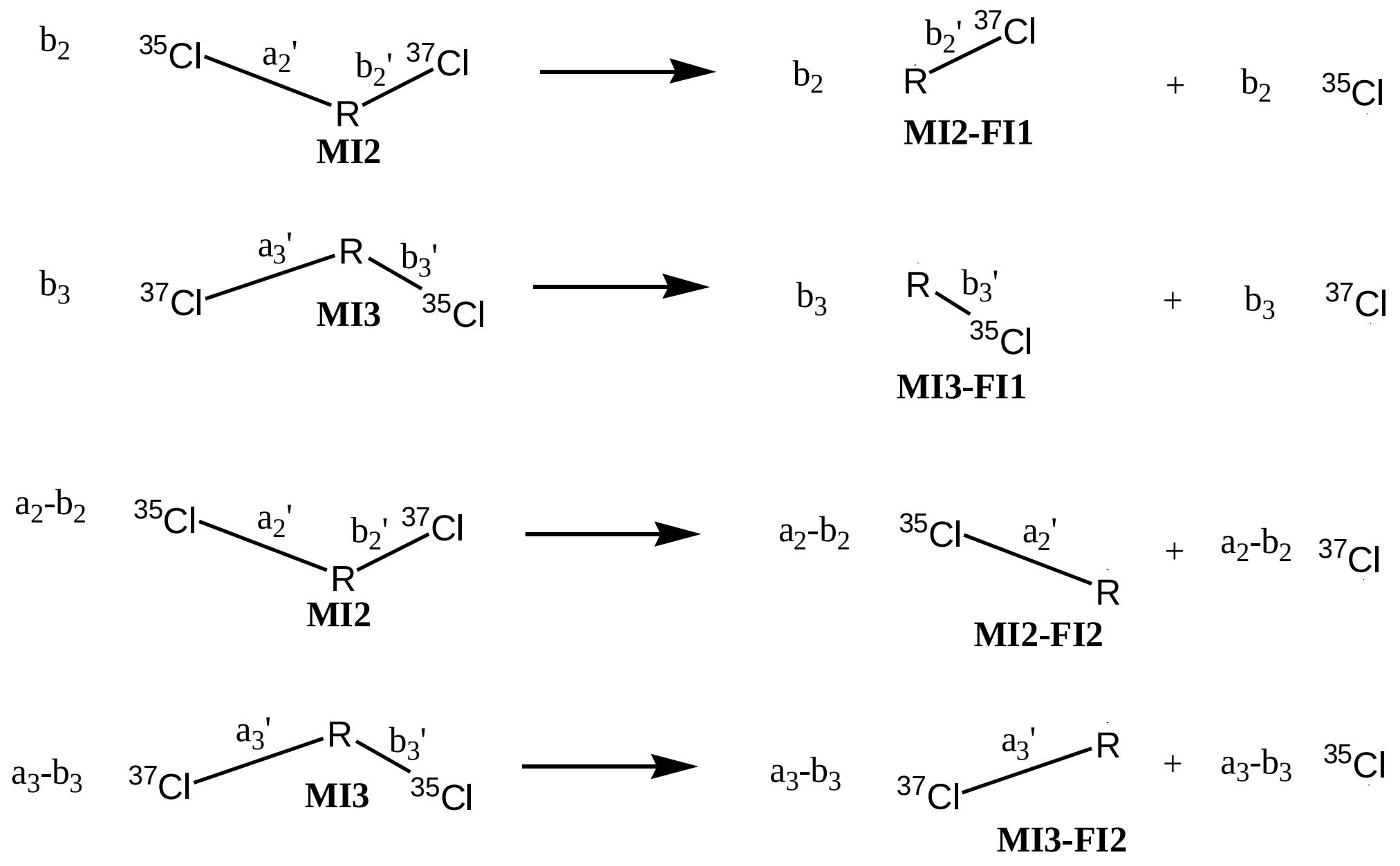

**Figure 3**. The four possible dehalogenation pathways of the MI2 and MI3 (molecular isotopomer 2 and 3) during fragmentation in EI-MS. The molar amounts of MI2 and MI3 reacting in the four dehalogenation pathways are $b_2$, $b_3$, $a_2 - b_2$, and $a_3 - b_3$, respectively. MI2-F1, MI3-F1, MI2-F2 and MI3-F2 are the four possible fragments generated from MI2 and/or MI3 through the four dehalogenation pathways.

The prerequisites for the occurrence of intramolecular isotope fractionation are 1) the isotopologue formula of a parent ion should possess both light and heavy isotopes; 2) at least two dehalogenation reactions can happen synchronously and alternatively at two reacting positions during the generation of the product ion from this parent ion; and 3) the isotopes on any reacting positions should not entirely leave from the parent ion. In addition, the reacting positions can be equivalent or non-equivalent in structure, provided the reactions taking place synchronously and alternatively.

As shown in Figure 3, the four possible fragments of MI2 and MI3 are named as MI2-F1, MI3-F1, MI2-F2 and MI3-F2, with molar amounts of $b_2$, $b_3$, $a_2 - b_2$, and $a_3 - b_3$, respectively.

In the light of quasiequilibrium theory (QET),[24] if the internal energy and time scale in EI-MS are fixed for a compound, then the fractions of all the ions stemming from this compound are certain values and predicable, indicating that the relative abundances of all the ions can reach an equilibrium state. Therefore, the equilibrium constant between the cleavages of bond $a_2$' and $a_3$' can be expressed as:

$$K_a = \frac{\frac{b_2}{a_2 - b_2}}{\frac{b_3}{a_3 - b_3}} = \frac{b_2(a_3 - b_3)}{b_3(a_2 - b_2)} \qquad (2.1)$$

And that between the cleavages of bond $b_3$' and $b_2$' can be expressed as:

$$K_b = \frac{\frac{a_3 - b_3}{b_3}}{\frac{a_2 - b_2}{b_2}} = \frac{b_2(a_3 - b_3)}{b_3(a_2 - b_2)} \qquad (2.2)$$

Therefore, $K_a = K_b$.

Similarly, the equilibrium constants ($K_2$ and $K_3$) between the cleavages of $a_2$' and $b_2$', and $a_3$' and $b_3$' can be expressed as:

$$K_2 = \frac{\frac{b_2}{a_2 - b_2}}{\frac{a_2 - b_2}{b_2}} = \left(\frac{b_2}{(a_2 - b_2)}\right)^2 \qquad (2.3)$$

$$K_3 = \frac{\frac{a_3 - b_3}{b_3}}{\frac{b_3}{a_3 - b_3}} = \left(\frac{a_3 - b_3}{b_3}\right)^2 \qquad (2.4)$$

The isotope ratio of the parent ion is

$$IR_{par} = \frac{a_2 + a_3}{a_2 + a_3} = 1 \qquad (2.5)$$

And that of the product ion is

$$IR_{pro(a_2a_3)} = \frac{b_2 + a_3 - b_3}{b_3 + a_2 - b_2} \qquad (2.6)$$

According to eqs 2.1-2.4, we have

$$b_2 = \frac{a_2}{1 + \frac{1}{\sqrt{K_2}}} \qquad (2.7)$$

$$b_3 = \frac{a_3}{1 + \sqrt{K_3}} \qquad (2.8)$$

$$K_a = \sqrt{K_2 K_3} \qquad (2.9)$$

Substitute eqs 2.7and 2.8 to eq 2.6, we have

$$IR_{pro(a_2a_3)} = \frac{\frac{a_2}{1+\frac{1}{\sqrt{K_2}}} + a_3 - \frac{a_3}{1+\sqrt{K_3}}}{\frac{a_3}{1+\sqrt{K_3}} + a_2 - \frac{a_2}{1+\frac{1}{\sqrt{K_2}}}} \qquad (2.10)$$

Substituting eq 2.9 into eq 2.10, gives

$$IR_{pro(a_2a_3)} = \frac{\frac{K_a}{K_a+\sqrt{K_3}} + \frac{a_3}{a_2}\frac{\sqrt{K_3}}{1+\sqrt{K_3}}}{\frac{a_3}{a_2}\frac{1}{1+\sqrt{K_3}} + \frac{\sqrt{K_3}}{K_a+\sqrt{K_3}}} \qquad (3.11)$$

Letting $\sqrt{K_3} = x$ and $\frac{a_3}{a_2} = R_{0(a_2a_3)}$, then we have

$$y=\frac{\dfrac{K_a}{K_a+x}+\dfrac{IR_{0(a_2a_3)}x}{1+x}}{\dfrac{IR_{0(a_2a_3)}}{1+x}+\dfrac{x}{K_a+x}} \qquad (2.12)$$

With differential calculation, eq 2.12 leads to

$$\frac{dy}{dx}=\frac{\left[\dfrac{-K_a}{(K_a+x)^2}+\dfrac{IR_{0(a_2a_3)}}{(1+x)^2}\right]\left(\dfrac{IR_{0(a_2a_3)}}{1+x}+\dfrac{x}{K_a+x}\right)-\left(\dfrac{K_a}{K_a+x}+\dfrac{IR_{0(a_2a_3)}}{1+x}\right)\left[\dfrac{-IR_{0(a_2a_3)}}{(1+x)^2}+\dfrac{K_a}{(K_a+x)^2}\right]}{\left(\dfrac{IR_{0(a_2a_3)}}{1+x}+\dfrac{x}{K_a+x}\right)^2} \qquad (2.13)$$

which simplifies to

$$\frac{dy}{dx}=\frac{(IR_{0(a_2a_3)}+1)\left[\dfrac{IR_{0(a_2a_3)}}{(1+x)^2}-\dfrac{K_a}{(K_a+x)^2}\right]}{\left(\dfrac{IR_{0(a_2a_3)}}{1+x}+\dfrac{x}{K_a+x}\right)^2} \qquad (2.14)$$

Because $a_2$ approximates to or equals $a_3$, letting $IR_{0-a_2a_3}=1$, then we have

$$y=\frac{\dfrac{K_a}{K_a+x}+\dfrac{x}{1+x}}{\dfrac{1}{1+x}+\dfrac{x}{K_a+x}} \qquad (2.15)$$

and

$$\frac{dy}{dx}=2\frac{(1-K_a)x^2+K_a{}^2-K_a}{\left(x^2+2x+K_a\right)^2} \qquad (2.16)$$

When $\frac{dy}{dx}=0$, we have $x=\pm\sqrt{K_a}$. Because $x\geq 0$, thus $x=\sqrt{K_a}$.

Because $K_a>1$, then the function $f(x)=(1-K_a)x^2+K_a{}^2-K_a$ has an inverted-U shape, and is greater than zero ( $f(x)>0$ ) at $x\in[0,\sqrt{K_a})$, and lower than zero ($f(x) < 0$) at $x \in (\sqrt{K_a}, \infty)$. Therefore, the function $f(x)$ is monotonically increasing at $x \in [0, \sqrt{K_a})$, and monotonically decreasing at $x \in (\sqrt{K_a}, \infty)$, and has a maximum at $x = \sqrt{K_a}$. When $K_3 = 0$, $IR_{pro(a_2a_3)}$ has the minimum:

$$IR_{pro(a_2a_3)} = 1 = IR_{0(a_2a_3)} \quad (2.17)$$

When $K_3 = K_a$, $IR_{pro(a_2a_3)}$ has the maximum:

$$IR_{pro(a_2a_3)} = \frac{x^2 + 2K_a x + K_a}{x^2 + 2x + K_a} = \frac{K_a + 2K_a\sqrt{K_a} + K_a}{K_a + 2\sqrt{K_a} + K_a} \quad (2.18)$$

$$IR_{pro(a_2a_3)} = \sqrt{K_a} \quad (2.19)$$

When $K_3$ approaches infinity, which results in

$$\lim_{x\to\infty} f(x) = \lim_{x\to\infty} \frac{x^2 + 2K_a x + K_a}{x^2 + 2x + K_a} = 1 \quad (2.20)$$

Then for $K_3 \in (0, \infty)$, we have

$$1 < IR_{pro(a_2a_3)} \le \sqrt{K_a} \quad (2.21)$$

Thus, the isotope ratio of a product ion is higher than that of its parent ion. The physicochemical meanings for the cases of $K_3 = 0$, $K_3 \to \infty$ and $K_3 = K_a$ are as follows.

$K_3 = 0$: the bonds $a_2$' and $a_3$' are completely broken, while the bonds $b_2$' and $b_3$' are completely unbroken. In this case, the isotope ratio of the product ion is equal to that of the parent ion, indicating no intramolecular isotope fractionation takes place. This phenomenon probably present for asymmetric molecules. If the critical energies of the asymmetric bonds are different sufficiently, so that a bond can be completely cleaved and another completely non-cleaved or with a negligible ratio cleaved.

$K_3 \to \infty$: this case is equivalent to the case of $K_3 = 0$.

$K_3 = K_a$: if the bonds a2', b2', a3' and b3' are structurally symmetric, then all these bonds are partially broken and the bonds linking to light isotope are more broken than those liking to the heavy, indicating the occurrence of intramolecular isotope fractionation.

For other cases, intramolecular isotope fractionation can take place, and the isotope ratio of the product ion is higher than 1 but lower than $\sqrt{K_a}$ .

**Impact on Apparent Isotope Ratio of a Product Ion**.

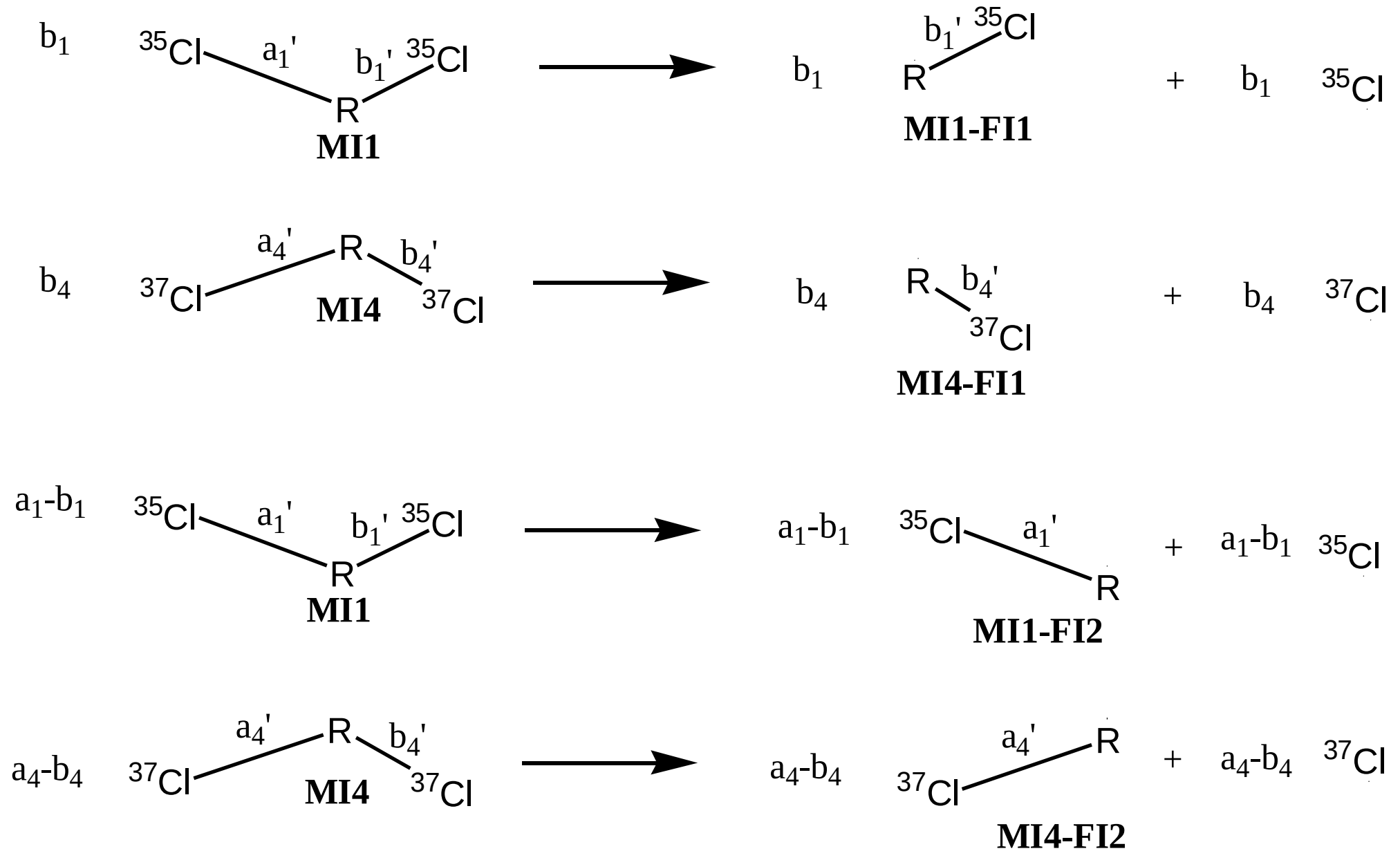

**Figure 4**. The four possible dehalogenation pathways of MI1 and MI4 (molecular isotopomer 1 and 4) during fragmentation in EI-MS. The molar amounts of MI1 and MI4 reacting in the four dehalogenation pathways are $b_1$, $b_4$, $a_1 - b_1$, and $a_4 - b_4$,

respectively. MI1-F1, MI4-F1, MI1-F2 and MI4-F2 are the four possible fragments generated from MI1 and/or MI4 through the four dehalogenation pathways.

According to the reaction pathways illustrated in Figure 3 and 4, the isotope ratio of total isotopologues of the product ion (apparent isotope ratio of the product ion) of the fictitious compound is

$$IR_{pro(app)} = \frac{b_2 + a_3 - b_3 + a_4}{b_3 + a_2 - b_2 + a_1} \qquad (2.22)$$

And the isotope ratio of the parent ion is

$$IR_{par} = \frac{a_2 + a_3 + 2a_4}{a_2 + a_3 + 2a_1} \qquad (2.23)$$

which transforms to

$$IR_{par} = \frac{\frac{a_2 + a_3}{2} + a_4}{\frac{a_2 + a_3}{2} + a_1} \qquad (2.24)$$

If the bonds a2’, b2’, a3’ and b3’ are structurally equivalent, then we have $a_2 = a_3$, and thus eq 2.24 can simplify to

$$IR_{par} = \frac{a_2 + a_4}{a_2 + a_1} \qquad (2.25)$$

As $IR_{pro(a_2 a_3)} = \sqrt{K_a}$ , we have

$$\frac{A_{^{37}Cl}}{A_{^{37}Cl + ^{35}Cl}} = \frac{\sqrt{K_a}}{\sqrt{K_a} + 1} \qquad (2.26)$$

$$\frac{A_{^{35}Cl}}{A_{^{37}Cl+^{35}Cl}} = \frac{1}{\sqrt{K_a}+1} \qquad (2.27)$$

Substituting eqs 2.26 and 2.27 into eq 2.22, gives

$$IR_{pro(app)} = \frac{2a_2 \frac{\sqrt{K_a}}{\sqrt{K_a}+1} + a_4}{2a_2 \frac{1}{\sqrt{K_a}+1} + a_1} \qquad (2.28)$$

Then, we compare the $IR_{pro(app)}$ with $IR_{par}$ .

Letting $K_a = 1$, we obtain

$$IR_{pro(app)} = \frac{2a_2 \frac{1}{1+1} + a_4}{2a_2 \frac{1}{1+1} + a_1} = \frac{a_2 + a_4}{a_2 + a_1} = IR_{par} \qquad (2.29)$$

And letting $\sqrt{K_a} = x$, we have

$$y = \frac{2a_2 \frac{x}{x+1} + a_4}{2a_2 \frac{1}{x+1} + a_1} \qquad (2.30)$$

With differential calculation, eq 2.30 gives

$$\frac{dy}{dx} = \frac{\frac{2a_2}{(x+1)^2}(a_1 + 2a_2 + a_4)}{(2a_2 \frac{1}{x+1} + a_1)^2} \qquad (2.31)$$

Then the function y is monotonically increasing in the whole definitional domain. Because the function $f(K_a) = \sqrt{K_a} = x$ is monotonically increasing in the whole definitional domain, then the function

$$y = g(x) = g[f(K_a)] = IR_{pro(app)} = \frac{2a_2 \dfrac{\sqrt{K_a}}{\sqrt{K_a}+1} + a_4}{2a_2 \dfrac{1}{\sqrt{K_a}+1} + a_1} \qquad (2.32)$$

is monotonically increasing.

When $K_a = 1$, we can get

$$IR_{pro(app)} = IR_{par} \qquad (2.33)$$

Therefore, when $K_a > 1$, we have

$$IR_{pro(app)} = \frac{2a_2 \dfrac{\sqrt{K_a}}{\sqrt{K_a}+1} + a_4}{2a_2 \dfrac{1}{\sqrt{K_a}+1} + a_1} > \frac{a_2 + a_4}{a_2 + a_1} = IR_{par} \qquad (2.34)$$

Thus, the apparent isotope ratio of the product ion is identically higher than that of the parent ion, only if $K_a > 1$.

For asymmetric molecules, letting $IR_{pro(a_2 a_3)} = K_u$, then we have $1 \le K_u < \sqrt{K_a}$. We do not take into account the case that the critical energies of the bonds linking to isotopes are sufficiently different so that one can be broken and another completely unbroken. In other words, the case of $K_u = 0$ is not taken into account. Then, we have $0 < K_u < \sqrt{K_a}$.

Substituting $IR_{pro(a_2 a_3)} = K_u$ into eq 2.22, leads to

$$IR_{pro(app)} = \frac{(a_2 + a_3)\dfrac{K_u}{K_u + 1} + a_4}{(a_2 + a_3)\dfrac{1}{K_u + 1} + a_1} \qquad (2.35)$$

Defining function

$$IR_{pro(app)} = y = f(K_u) = \frac{(a_2 + a_3)\dfrac{K_u}{K_u + 1} + a_4}{(a_2 + a_3)\dfrac{1}{K_u + 1} + a_1} \qquad (2.36)$$

and performing differential calculation, result in

$$\frac{dy}{dK_u} = \frac{(a_2 + a_3)(a_1 + a_2 + a_3 + a_4)}{(K_u + 1)^2[(a_2 + a_3)\dfrac{1}{K_u + 1} + a_1]^2} = \frac{(a_2 + a_3)(a_1 + a_2 + a_3 + a_4)}{[a_2 + a_3 + (K_u + 1)a_1]^2} \qquad (2.37)$$

Therefore $\frac{dy}{dK_u} > 0$, and the function y is monotonically increasing.

As

$$K_u = \frac{x^2 + 2K_a x + K_a}{x^2 + 2x + K_a}, (x = \sqrt{K_3}) \qquad (2.38)$$

when $0 \le x \le \sqrt{K_a}$, the function $K_u(x)$ is monotonically increasing, thus the function $y = f(K_u) = f[K_u(x)]$ is monotonically increasing.

When $x \ge \sqrt{K_a}$, the function $K_u(x)$ is monotonically decreasing, thus the function $y = f(K_u) = f[K_u(x)]$ is monotonically decreasing.

When, $x = \sqrt{K_a}$, the function y has the maximum:

$$y = \frac{(a_2 + a_3)\frac{\sqrt{K_a}}{\sqrt{K_a}+1} + a_4}{(a_2 + a_3)\frac{1}{\sqrt{K_a}+1} + a_1} \qquad (2.39)$$

When $x = 0$, then $K_u = IR_{pro(a_2a_3)} = 1 = IR_{0(a_2a_3)}$, and the function has the minimum:

$$y = \frac{(a_2 + a_3)\frac{1}{1+1} + a_4}{(a_2 + a_3)\frac{1}{1+1} + a_1} = \frac{\frac{a_2 + a_3}{2} + a_4}{\frac{a_2 + a_3}{2} + a_1} = IR_{par} \qquad (2.40)$$

Namely, $IR_{pro(app)} = IR_{par}$.

When $x \to \infty$, then the limit of $K_u$ can be given as

$$\lim_{x\to\infty} K_u = \lim_{x\to\infty} \frac{x^2 + 2K_a x + K_a}{x^2 + 2x + K_a} = 1 \qquad (2.41)$$

Hence, the limit of the function y is

$$\lim_{K_u\to 1} y = \lim_{K_u\to 1} \frac{(a_2 + a_3)\frac{K_u}{K_u+1} + a_4}{(a_2 + a_3)\frac{1}{K_u+1} + a_1} = \frac{(a_2 + a_3)\frac{1}{1+1} + a_4}{(a_2 + a_3)\frac{1}{1+1} + a_1} = \frac{\frac{a_2 + a_3}{2} + a_4}{\frac{a_2 + a_3}{2} + a_1} \qquad (2.42)$$

Namely, $IR_{pro(app)} = IR_{par}$.

For general cases, as

$$K_u = R_{pro(a_2a_3)} = \frac{x^2 + 2K_a x + K_a}{x^2 + 2x + K_a} \qquad (2.43)$$

Hence, the $IR_{pro(app)}$ can be expressed as

$$IR_{pro(app)} = \frac{(a_2+a_3)\dfrac{\dfrac{x^2+2K_a x+K_a}{x^2+2x+K_a}}{\dfrac{x^2+2K_a x+K_a}{x^2+2x+K_a}+1}+a_4}{(a_2+a_3)\dfrac{1}{\dfrac{x^2+2K_a x+K_a}{x^2+2x+K_a}+1}+a_1}\ ,\ (x=\sqrt{K_3}) \quad (2.44)$$

For $x \in (0, \infty)$, we have the following inequation

$$IR_{par} < IR_{pro(app)} \le \frac{(a_2+a_3)\dfrac{\sqrt{K_a}}{\sqrt{K_a}+1}+a_4}{(a_2+a_3)\dfrac{1}{\sqrt{K_a}+1}+a_1} \qquad (2.45)$$

Thus, the apparent isotope ratio of a product ion is always higher than that of its parent ion, indicating normal isotope fractionation for the product ion occurring in EI-MS. Since intermolecular isotope fractionation is taken into consideration in eq 2.44, thus normal isotope fractionation for the product ion can take place although the inverse intermolecular fractionation presents during the dehalogenation from the parent ion to the product ion, provided that the intramolecular isotope fractionation exists.

# PART 3: EFFECTS TO ISOTOPOLOGUE DISTIBUTION AND IMPLICATIONS TO ISOTOPE RATIO EVALUATION SCHEMES

## Case of Symmetric Compounds.

*Perspective from the Non-Dehalogenated (Detected) Molecular Ion.*

Presently, the isotope ratio evaluation schemes used for CSIA-Cl/Br studies using GC-EI-MS are based on the binomial theorem. Isotope ratio (R) can be calculated with any pair of neighboring isotopologues by the isotopologue-pair scheme: [Elsner, 2008; Jin 2011]

$$IR = \frac{i}{n-i+1} \cdot \frac{I_i}{I_{i-1}} \qquad (3.1)$$

Where IR is the isotope ratio ($^{37}Cl/^{35}Cl$ or $^{81}Br/^{79}Br$), n is the number of the Cl/Br atoms of a certain ion, *i* is the number of the heavy isotope atoms ($^{37}Cl$/ or $^{81}Br$) in a specific ion isotopologue, and *I* is the measured abundance (mass spectrometric intensity) of the ion isotopologue. The prerequisite for this evaluation scheme is that the measured abundances of the ion isotopologues comply with binomial distribution.

In a previous study, we developed an isotope ratio evaluation scheme using complete isotopologues. This complete-isotopologue scheme can be expressed as fowllows:

$$IR_{Comp_Iso} = \frac{\sum_{i=0}^{n} iI_i}{\sum_{i=0}^{n} (n-i)I_i} \qquad (3.2)$$

where $IR_{Comp_Iso}$ is the isotope ratio calculated with the complete-isotopologue scheme. If the isotopolohues indeed conform to binomial distribution, the isotope ratio calculated with the isotopologue-pair scheme must equal that calculated by the complete-isotopologue scheme, namely,

$$\frac{i}{n-i+1}\cdot\frac{I_i}{I_{i-1}}=\frac{\sum_{i=0}^{n} iI_i}{\sum_{i=0}^{n}(n-i)I_i} \qquad (3.3)$$

We hypothesize the ionized molecular ion isotopologues of an organochlorine compound comply with binomial distribution. Actually, if the chlorination processes for synthesizing this compound are pseudo-first order reactions (Cl excessive) and the reacting positions are structurally symmetric, then the chlorine isotope ratios on the reacting positions are identical, and accordingly the abundances of chlorine isotopologues of the compound comply with binomial distribution. In addition, we hypothesize that isotope fractionation cannot take place during the ionization process (not containing fragmentation processes) in EI-MS, which is theoretically reasonable. Therefore, the total molecular ions (containing both fragmented and non-fragmented) and the initial molecules of the compound have the same distribution of chlorine isotopologues (binomial distribution). The isotope ratio of the total molecular ions thus can be calculated as:

$$IR_0=\frac{i}{n-i+1}\cdot\frac{I_{i0}}{I_{(i-1)0}} \qquad (3.4)$$

Where $IR_0$ is the isotope ratio of the total molecular ions. The non-fragmented molecular ion is actually the detected molecular ion, while the fragmented molecular ion cannot be observed.

We hypothesize that the ratio of the detected molecular ion (non-fragmented) relative to the total molecular ions with the same isotopologue formula is:

$$a_1 = \frac{I_{1-1}}{I_{(1-1)0}}, a_2 = \frac{I_{2-1}}{I_{(2-1)0}} \ldots a_i = \frac{I_{i-1}}{I_{(i-1)0}} \ldots a_n = \frac{I_{n-1}}{I_{(n-1)0}}, a_{n+1} = \frac{I_n}{I_{n0}} \qquad (3.5)$$

Then for a random pair of neighboring isotopologue ions with *i*-1and *i* heavy isotope atoms, respectively, according to eqs 3.1 and 3.5, we have

$$IR = \frac{a_{i+1} I_{i0}}{a_i I_{(i-1)0}} \cdot \frac{i}{n-i+1} \qquad (3.6)$$

which transforms to

$$\frac{I_{i0}}{I_{(i-1)0}} \cdot \frac{i}{n-i+1} = \frac{a_i}{a_{i+1}} IR = IR_0 \qquad (3.7)$$

Similarly, for the pair of isotopologue ions with *I* and *i*+1 heavy isotope atoms, respectively, we obtain

$$\frac{I_{(i+1)0}}{I_{i0}} \cdot \frac{i+1}{n-i} = \frac{a_{i+1}}{a_{i+2}} IR = IR_0 \qquad (3.8)$$

If the detected molecular ion isotopologues comply with binomial distribution, then the isotope ratios calculated with random different pairs of neighboring isotopologue ions are equal. Thus, for the isotope ratios calculated with random three neighboring isotopologue ions, we have

$$\frac{a_i}{a_{i+1}} IR = \frac{a_{i+1}}{a_{i+1+1}} IR \qquad (3.9)$$

which simplifies to

$$\frac{a_i}{a_{i+1}} = \frac{a_{i+1}}{a_{i+1+1}} \qquad (3.10)$$

It can be seen from eq 3.10 that the progression ($a_1$, $a_2$ … $a_i$, $a_{i+1}$ … $a_n$, $a_{n+1}$) is geometric:

$$a_i = a_1 q^{i-1} \qquad (3.11)$$

of which the common ratio (q) is

$$q = \frac{a_2}{a_1} = \frac{a_{i+1}}{a_i} \qquad (3.12)$$

Therefore, if the detected molecular ion isotopologues comply with binomial distribution, then the progression ($a_1$, $a_2$ … $a_i$, $a_{i+1}$ … $a_n$, $a_{n+1}$) must be geometric.

Due to the presence of intermolecular isotope fractionation, the isotopologues containing less heavy-isotope atoms are more liable to be dehalogenated relative to those containing more heavy-isotope atoms, in other words, the heavier isotopeloues are more prone to remain in the non-fragmented ions than the lighter ones. For the detected molecular ion, the ratio ($a_i$) for a lighter isotopologue ion is lower than that for a heavier one ($a_{i+1}$). Thus, q is higher than 1 ($q > 1$).

If $n \to \infty$, the limit of $a_i$ is

$$\lim_{i \to n} a_i = \lim_{i \to n} a_1 q^{i-1} = \infty > 1 \qquad (3.13)$$

This contradicts the reality, because the abundance of the detected molecular ion is impossibly higher than that of the total molecular ions. Therefore, in reality, the progression ($a_1$, $a_2$ … $a_i$, $a_{i+1}$ … $a_n$, $a_{n+1}$) should never be geometric. We thus deduce that the detected molecular ion isotopologues do not comply with binomial distribution.

Theoretically, the maximum of $a_{n+1}$ is can only approach to 1, and that is

$$\lim_{n \to \infty} a_{n+1} = \lim_{n \to \infty} \frac{I_n}{I_{n0}} = 1 \qquad (3.14)$$

which is also unlikely in reality, due to that the heavy isotopologues cannot be non-dechlorinated in at all EI-MS.

Thus, the progression ($a_1$, $a_2$ … $a_i$, $a_{i+1}$ … $a_n$, $a_{n+1}$) should be deceleratingly incremental, namely, the ratios ( $a_{i+1} / a_i = q_i$ ) decrease gradually. Hence, we receive

$$\frac{a_{i+1}}{a_i} > \frac{a_{i+2}}{a_{i+1}} \qquad (3.15)$$

According to eqs 3.7 and 3.8, we have

$$IR_i = \frac{a_{i+1}}{a_i} IR_0 \qquad (3.16)$$

$$IR_{i+1} = \frac{a_{i+2}}{a_{i+1}} IR_0 \qquad (3.17)$$

Therefore, $IR_i > IR_{i+1}$ can be obtained.

The isotope ratio ($IR_1$) calculated with $I_0$ and $I_1$ is the highest among those calculated with any possible pair of neighboring isotopologues, and higher than that calculated with complete molecular ion isotopologues by the complete-isotopologue scheme of isotope ratio evaluation [Tang et al, 2017].

*Perspective from the Dehalogenated (Unobserved) Molecular Ion.*

If the dehalogenated molecular ion isotopologues comply with binomial distribution, then we get

$$IR' = \frac{i}{n-i+1} \cdot \frac{I'_i}{I'_{i-1}} \qquad (3.18)$$

where IR' is the isotope ratio of the dehalogenated molecular ion, *I'* is the abundance of the dehalogenated molecular ion.

The ratio of the dehalogenated molecular ion relative to the total molecular ions is hypothesized as

$$a'_1 = \frac{I'_{1-1}}{I'_{(1-1)0}}, a'_2 = \frac{I'_{2-1}}{I'_{(2-1)0}} ... a'_i = \frac{I'_{i-1}}{I'_{(i-1)0}} ... a'_n = \frac{I'_{n-1}}{I'_{(n-1)0}}, a'_{n+1} = \frac{I'_n}{I'_{n0}} \qquad (3.19)$$

According to eqs 3.18 and 3.19, the isotope ratio (IR') calculated with a random pair of neighboring isotopologues (i-1 and i) is

$$IR' = \frac{a'_{i+1} I'_{i0}}{a'_i I'_{(i-1)0}} \cdot \frac{i}{n-i+1} \qquad (3.20)$$

which transforms to

$$\frac{I'_{i0}}{I'_{(i-1)0}} \cdot \frac{i}{n-i+1} = \frac{a'_i}{a'_{i+1}} IR' = IR_0 \qquad (3.21)$$

Similarly for the pair of isotopologue *i* and *i*+1, we obtain

$$\frac{I'_{(i+1)0}}{I'_{i0}} \cdot \frac{i+1}{n-i} = \frac{a'_{i+1}}{a'_{i+2}} IR' = IR_0 \qquad (3.22)$$

If the isotopologues of the dehalogenated molecular ion comply with binomial distribution, then the isotope ratios calculated using random different pairs of neighboring isotopologues are equal. Therefore, for any three adjacent isotopologues, we have

$$\frac{a'_i}{a'_{i+1}} \cdot IR' = \frac{a'_{i+1}}{a'_{i+1+1}} IR' \qquad (3.23)$$

which simplifies to

$$\frac{a'_i}{a'_{i+1}} = \frac{a'_{i+1}}{a'_{i+1+1}} \qquad (3.24)$$

Hence, the progression (a'$_1$, a'$_2$ … a'$_i$, a'$_{i+1}$ … a'$_n$, a'$_{n+1}$) is geometric:

$$a'_i = a'_1 q'^{i-1} \qquad (3.25)$$

of which the common ratio (q') is

$$q' = \frac{a'_2}{a'_1} = \frac{a'_{i+1}}{a'_i} \qquad (3.26)$$

Due to intermolecular isotope fractionation, the lighter molecular isotopologues are more liable to be dehalogenated compared with the heavier ones, the common ratio is thus less than 1 ($q' < 1$).

We hypothesize $n \to \infty$, then get the limit of a'$_i$

$$\lim_{i \to n} a'_i = \lim_{i \to n} a'_1 q'^{i-1} = 0 \qquad (3.27)$$

which contradicts the reality, because it is impossible that the heavier isotopologues are completely non-dehalogenated. Therefore, in fact, the progression (a'$_1$, a'$_2$ … a'$_i$, a'$_{i+1}$ … a'$_n$, a'$_{n+1}$) is non-geometric. We therefore conclude that the dehalogenated molecular ion isotopologues do not comply with binomial distribution neither.

**Case of Asymmetric Compounds.**

We hypothesize that the isotopologues of an asymmetric organochlorine compound comply with binomial distribution which can be expressed as

$$f_0(n) = (a_0 + b_0)^n \qquad (3.28)$$

in which $a_0$ and $b_0$ denote the relative abundances of the light and the heavy isotopes, respectively, and have the following relationships:

$$\frac{b_0}{a_0} = IR_0 \qquad (3.29)$$

$$a_0 + b_0 = 1 \qquad (3.30)$$

If this compound lost the Cl atoms from only one specified position during fragmentation in EI-MS, then the chlorine isotope ratio on this specified position of the detected molecular ion become higher. However, the chlorine isotope ratios on the rest positions are unchanged. If the Cl on the specified position is not taken into account, then the isotopologues comply with binomial distribution. When taking into consideration of the Cl on the specified position, we can express the distribution of the isotopologues as

$$f(n) = (a + b)(a_0 + b_0)^{n-1} \qquad (3.31)$$

where a and b represent the ratios of $^{35}Cl/(^{35}Cl+^{37}Cl)$ and $^{37}Cl/(^{35}Cl+^{37}Cl)$ on the specified position, respectively.

Because $(a_0 + b_0)^{n-1}$ is a binomial expression and $b/a > b_0/a_0$, the function $f(n)$ thus does not comply with binomial distribution. Therefore, for the asymmetric compound, the detected molecular ion isotopologues do not obey binomial distribution. However, the total dehalogenated product ions (prior to next-step dehalogenation process) of this compound still obey binomial distribution. Nevertheless, the isotopologues of the detected product ion do not obey binomial distribution due to the occurrence of intermolecular isotope fractionation during the next-step dehalogenation, no matter of whether the remaining Cl atoms are symmetric or not.

We list three random adjacent pairs of combinable similar terms of $f\left(n\right)$ as follows.

$$bp_{i-1} = bC_{n-1}^{i-1} a_0^{\ n-1-(i-1)} b_0^{\ i-1} \qquad (3.32)$$

$$ap_i = aC_{n-1}^{i} a_0^{\ n-1-i} b_0^{\ i} \qquad (3.33)$$

$$bp_i = bC_{n-1}^{i} a_0^{\ n-1-i} b_0^{\ i} \qquad (3.34)$$

$$ap_{i+1} = aC_{n-1}^{i+1} a_0^{\ n-1-(i+1)} b_0^{\ i+1} \qquad (3.35)$$

$$bp_{i+1} = bC_{n-1}^{i+1} a_0^{\ n-1-(i+1)} b_0^{\ i+1} \qquad (3.36)$$

$$ap_{i+2} = aC_{n-1}^{i+2} a_0^{\ n-1-(i+2)} b_0^{\ i+2} \qquad (3.37)$$

Then the isotope ratios ($IR_{i+1}$ and $IR_{i+2}$) calculated with the pairs of neighboring isotopologues are

$$IR_{i+1} = \frac{bp_i + ap_{i+1}}{bp_{i-1} + ap_i} \cdot \frac{i+1}{n-i} = \frac{bC_{n-1}^{i} a_0^{\ n-1-i} b_0^{\ i} + aC_{n-1}^{i+1} a_0^{\ n-1-(i+1)} b_0^{\ i+1}}{bC_{n-1}^{i-1} a_0^{\ n-1-(i-1)} b_0^{\ i-1} + aC_{n-1}^{i} a_0^{\ n-1-i} b_0^{\ i}} \cdot \frac{i+1}{n-i} \qquad (3.38)$$

$$IR_{i+2} = \frac{bp_{i+1} + ap_{i+2}}{bp_i + ap_{i+1}} \cdot \frac{i+2}{n-i-1} = \frac{bC_{n-1}^{i+1} a_0^{\ n-1-(i+1)} b_0^{\ i+1} + aC_{n-1}^{i+2} a_0^{\ n-1-(i+2)} b_0^{\ i+2}}{bC_{n-1}^{i} a_0^{\ n-1-i} b_0^{\ i} + aC_{n-1}^{i+1} a_0^{\ n-1-(i+1)} b_0^{\ i+1}} \cdot \frac{i+2}{n-i-1} \qquad (3.39)$$

which simplify and transform to

$$\frac{bp_i + ap_{i+1}}{bp_{i-1} + ap_i} \cdot \frac{i+1}{n-i} = \frac{\dfrac{ba_0 b_0}{i(n-1-i)} + \dfrac{ab_0^{\ 2}}{i(i+1)}}{\dfrac{ba_0^{\ 2}}{(n-i)(n-i-1)} + \dfrac{aa_0 b_0}{i(n-1-i)}} \cdot \frac{i+1}{n-i} \qquad (3.40)$$

$$\frac{bp_{i+1}+ap_{i+2}}{bp_i+ap_{i+1}}\cdot\frac{i+2}{n-i-1}=\frac{\frac{ba_0b_0}{(i+1)(n-1-i-1)}+\frac{ab_0^{\ 2}}{(i+2)(i+1)}}{\frac{ba_0^{\ 2}}{(n-i-1)(n-1-i-1)}+\frac{aa_0b_0}{(i+1)(n-1-i-1)}}\cdot\frac{i+2}{n-i-1} \qquad (3.41)$$

Then the proof of $IR_{i+1} > IR_{i+2}$ is equivalent to

$$\frac{bp_i+ap_{i+1}}{bp_{i-1}+ap_i}\cdot\frac{i+1}{n-i}>\frac{bp_{i+1}+ap_{i+2}}{bp_i+ap_{i+1}}\cdot\frac{i+2}{n-i-1} \qquad (3.42)$$

which is further equivalent to

$$\frac{\frac{ba_0}{i(n-1-i)}+\frac{ab_0}{i(i+1)}}{\frac{ba_0}{(n-i)(n-i-1)}+\frac{ab_0}{i(n-1-i)}}\cdot\frac{i+1}{n-i}>\frac{\frac{ba_0}{(i+1)(n-1-i-1)}+\frac{ab_0}{(i+2)(i+1)}}{\frac{ba_0}{(n-i-1)(n-1-i-1)}+\frac{ab_0}{(i+1)(n-1-i-1)}}\cdot\frac{i+2}{n-i-1} \quad (3.43)$$

Letting $\frac{ba_0}{ab_0}=k$ and substituting it into ineq 3.43, yield

$$\frac{\frac{k}{i(n-1-i)}+\frac{1}{i(i+1)}}{\frac{k}{(n-i)(n-i-1)}+\frac{1}{i(n-1-i)}}\cdot\frac{i+1}{n-i}>\frac{\frac{k}{(i+1)(n-1-i-1)}+\frac{1}{(i+2)(i+1)}}{\frac{k}{(n-i-1)(n-1-i-1)}+\frac{1}{(i+1)(n-1-i-1)}}\cdot\frac{i+2}{n-i-1} \quad (3.44)$$

which simplifies to

$$\frac{k(i+1)+(n-i-1)}{ki+(n-i)}>\frac{k(i+2)+(n-i-2)}{k(i+1)+(n-i-1)} \qquad (3.45)$$

This inequation can be transformed to

$$\frac{k(i+1)+(n-i-1)}{ki+(n-i)}>\frac{k[(i+1)+1]+[n-(i+1)-1]}{k(i+1)+[n-(i+1)]} \qquad (3.46)$$

We hypothesize a function $f(i)$ as

$$f(i) = \frac{k(i+1) + (n-i-1)}{ki + (n-i)} \qquad (3.47)$$

With differential calculation, we have

$$\frac{df(i)}{di} = \frac{(k-1)(ki-i+n) - (ki-i+k+n-1)(k-1)}{[ki+(n-i)]^2} = \frac{-(k-1)^2}{[ki+(n-i)]^2} \qquad (3.48)$$

When $k = 1$, we obtain $\frac{df(i)}{di} = 0$ which indicates that $f(i)$ identically equals a constant ( $f(i) = 1$ ). Thus

$$\frac{k(i+1) + (n-i-1)}{ki + (n-i)} = \frac{k(i+2) + (n-i-2)}{k(i+1) + (n-i-1)} = 1 \qquad (3.49)$$

and

$$IR_{i+1} = IR_{i+2} \qquad (3.50)$$

Due to intermolecular isotope fractionation, the isotope ratio of the remaining Cl atoms on the specified position increases, therefore we have

$$\frac{b}{a} > \frac{b_0}{a_0} \qquad (3.51)$$

which leads to

$$k = \frac{ba_0}{ab_0} > 1 \qquad (3.52)$$

Hence, we have

$$\frac{df(i)}{di} < 0 \qquad (3.53)$$

Accordingly, the function $f(i)$ monotonically decreases in the definitional domain. Therefore, the following inequation $f(i) > f(i+1)$ is obtained, namely,

$$\frac{k(i+1)+(n-i-1)}{ki+(n-i)} > \frac{k[(i+1)+1]+[n-(i+1)-1]}{k(i+1)+[n-(i+1)]} \qquad (3.54)$$

which is equivalent to

$$IR_{i+1} > IR_{i+2} \quad (i=0,1,2...n) \qquad (3.55)$$

Accordingly, in terms of asymmetric compounds, the isotope ratios calculated with the pairs of neighbouring isotopologues monotonically decrease with the increase of heavy-isotope atoms. As a result, the isotope ratio ($IR_1$) of the molecular ion of a compound calculated with the first pair of neighboring isotopologues ($I_0$ and $I_1$) is the largest, and higher than that calculated by the complete-isotopologue scheme using all the molecular isotopologues.

**Effects of Initial Molecular Isotopologues Distribution.**

The theoretical deduction provided above is based on the prerequisite that the molecular isotopologues of a compound comply with binomial distribution. If the initial molecular isotopologue distribution is not binomial, how does it affect the distribution of the detected molecular ion isotopologues? Obviously, if the initial molecular isotopologues do not comply with binomial distribution, the detected molecular ion isotopologues are extremely unlikely to comply with binomial distribution, and the former to some extent will possibly impact the latter. Thus, the isotope ratios calculated with isotopologue-pair scheme may not gradually decrease with the increase of the number of heavy isotope atoms, and that calculated with the first pair of neighboring isotopologues may be not the largest. Certainly, the isotope ratio calculated with the first pair of neighboring isotopologues cannot reflect that of the total isotopologues.

**Concerns for Product Ions.**

In addition, for a product ion of a compound, due to the presence of further intermolecular isotope fractionation during the next-step dehalogenation process, the isotopologues of the detected product ion thus never obey binomial distribution, which is similar to the molecular-ion situation discussed above. Similarly, the isotope ratios calculated with the pairs of neighbouring isotopologues of the product ion may gradually decrease with the increase of heavy-isotope atoms. Therefore, the isotope ratio ($IR_1$) of the product ion calculated with the first pair of neighbouring isotopologues ($I_0$ and $I_1$) may be the largest, and anticipated to exceed that calculated by the complete isotopologue-ion scheme using all the isotopologues of the product ion. However, if the isotopologues of the total product ions (further dehalogenated and non-dehalogenated) do not obey binomial distribution, then the isotope ratios calculated with the pairs of neighboring isotopologues of the product ion may not gradually decrease with the increase of heavy-isotope atoms, but be more susceptible to the initial distribution of the isotopologues of the total product ions. In addition, if intramolecular isotope fractionation presents in a dehalogenation from a parent ion to the corresponding product ion, then the isotopologues of the detected product ion trend to more unlikely comply with binomial distribution. This may result in that the observed distribution of the product ion isotopologues does not conform to the theoretical deduction that the isotope ratios calculated with the pairs of neighboring isotopologues of the product ion gradually decrease with the increase of heavy-isotope atoms.

# PART 4: IMPLICATIONS TO CSIA-Cl/Br STUDY

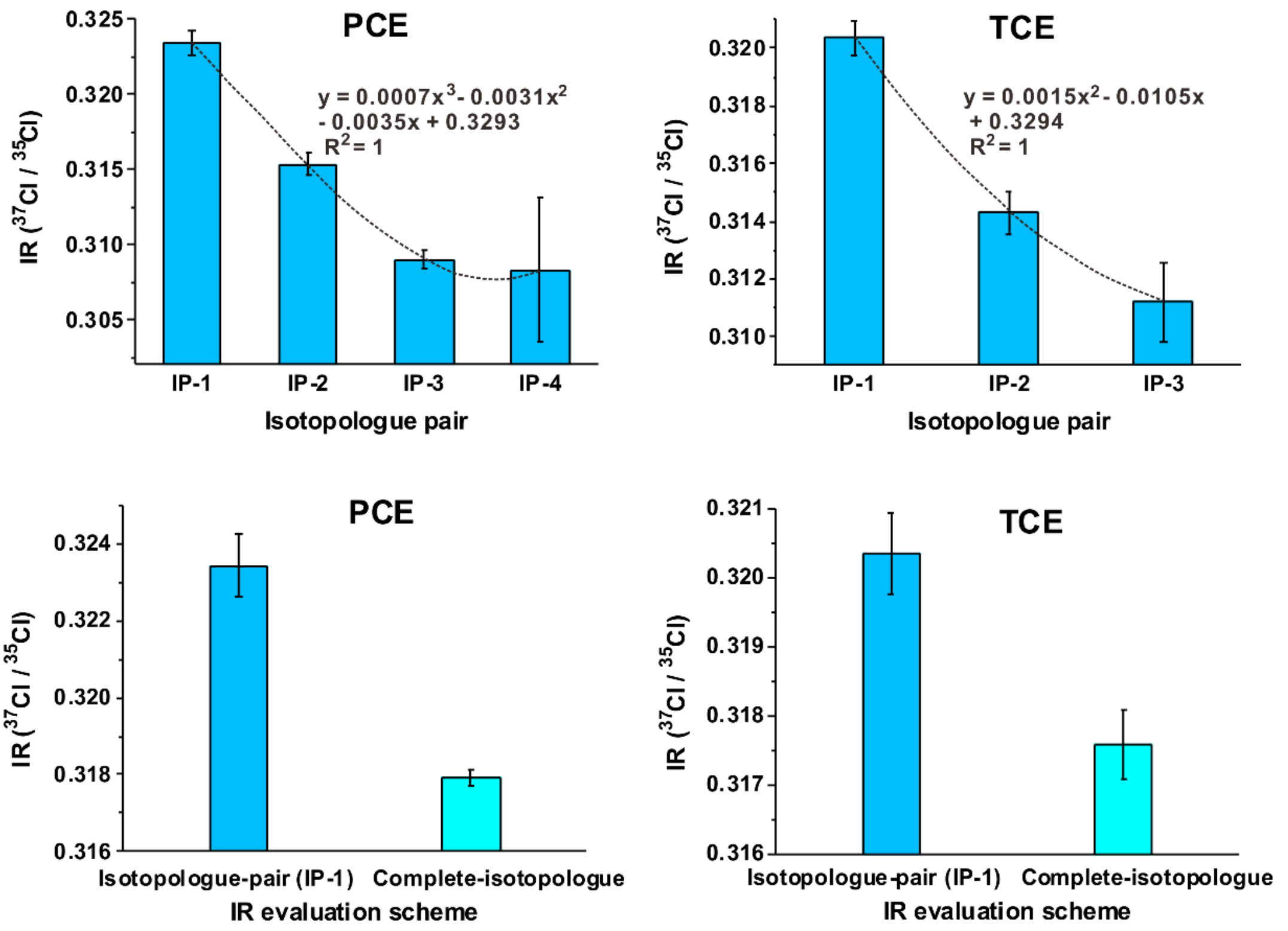

**Figure 5**. Measured chlorine isotope ratios of perchloroethylene (PCE) and trichloroethylene (TCE) calculated with the isotopologue-pair scheme using different pairs of neighboring molecular isotopologues, and those calculated with the complete-isotopologue scheme. IP: isotopologue pair, IR: isotope ratio. The correlations between the measured chlorine isotope ratios (y) and the numbers of heavy isotope atoms (x) are fitted with polynomial functions. Error bars show the standard deviations (1 σ). The standards PCE and TCE (high performance liquid chromatography grade) were bought from the manufacturer-1. The injection replicates for PCE were five, and six for TCE.

Because of the intermolecular isotope fractionation taking place in the dehalogenation processes in EI-MS, the detected molecular-ion and product-ion isotopologues never obey binomial distribution, and may only approximately comply with binomial distribution.

Up to now, we have not found any organochlorine compound whose molecular-ion and product-ion isotopologues strictly complied with binomial distribution, which is well in line with the theoretical conclusions in this study. On the other hand, the distributions of the molecular-ion and product-ion isotopologues of organobromines were more close to binomial distribution compared with those of organochlorines. This observation might ascribe to the lower isotope fractionation extents in EI-MS and less precise measurement results of organobromines in comparison with those of organochlorines (details will be provided in another incoming manuscript).

As indicated in Figure 5, the measured chlorine isotope ratios using different isotopologue pairs or different evaluation schemes agreed well with the theoretically inferences in this study. The isotope ratios of the molecular ions of PCE and TCE calculated with the pairs of neighboring isotopologues gradually decreased as the number of Cl atoms of the isotopologues increased, and the reduction was decelerated gradually. In addition, just as anticipated by the theoretical deduction, the isotope ratios of each compound calculated with the first pair of neighboring isotopologues are significantly higher than those calculated with the complete-isotopologue scheme (Figure 5).

Therefore, although external isotopic standards are utilized in CISA-Cl using GC-EI-MS, if the distribution modes of a target anlyte and the corresponding external isotopic standard(s) are not consistent, then the measured isotope ratios calculated by the isotopologue-pair scheme may be inaccurate. This inference has been experimentally proved in our laboratory by using the PCE and TCE from different manufacturers for CISA-Cl (Figure 6).

As indicated in Figure 6, the isotope ratios of PCE and TCE from the manufacturer-1 calculated with the corresponding first pair of neighboring molecular-ion isotopologues were higher than those from the manufacturer-2. However, the isotope ratios of the two compounds from the manufacturer-1 calculated by the complete-isotopologue scheme using the total molecular ion isotopologues were lower than those from the manufacturer-2.

The isotope ratios calculated with the complete-isotopologue scheme certainly reflected the comprehensive isotope ratios of the detected molecular ions of the compounds, showing real relative isotope ratios before calibration with external isotopic standards. While the isotope ratios calculated by isotopologue-pair scheme only reflected the idealized isotope ratios with the prerequisite that the molecular ion isotopologues of each compound obeyed binomial distribution. We hypothesize that the real isotope ratios (referenced to SMOC) of the two compounds from manufacturer-1 are known, then can use them as the external isotopic standards for those from manufacturer-2. Therefore, the real isotope ratios of the compounds from manufacturer-2 can be obtained by referencing to those from manufacturer-1. However, the real isotope ratios of the compounds from manufacturer-2 calculated with the isotopologue-pair scheme should be significantly lower than those calculated by the complete-isotopologue scheme (Figure 6). As a result, the real isotope ratios of the compounds from manufacturer-2 calculated with isotopologue-pair scheme cannot exactly indicate the isotope ratios in reality. Using external isotopic standards along with the complete molecular-isotopologue scheme can help to obtain the measured real isotope ratios accurately reflecting the reality.

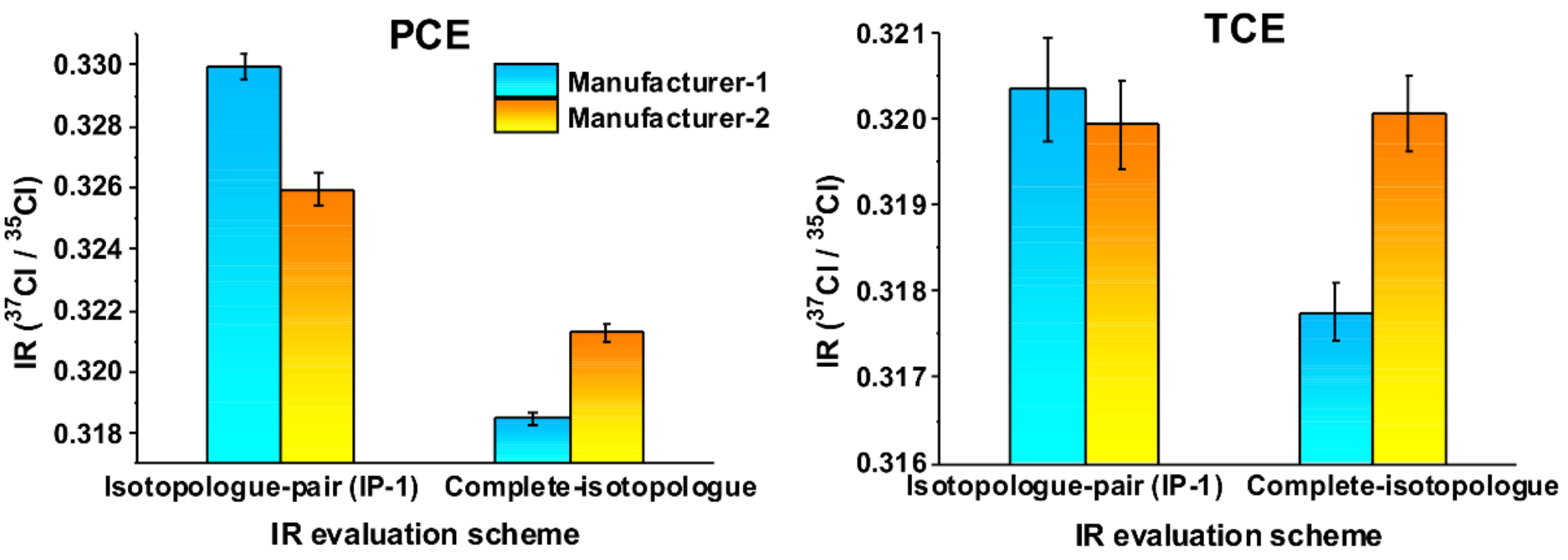

**Figure 6**. Measured chlorine isotope ratios of PCE and TCE from two different manufacturers calculated with the isotopologue-pair scheme using the first pair of neighboring molecular isotopologues, and those calculated with the complete-isotopologue scheme. The standards PCE and TCE purchased from the manufacturer-2 were of analytical reagent grade. The standards from different manufacturers were analyzed alternately and successively, and the injection replicates were six.

## CONCLUSIONS

In this study, a solid theoretical deduction concerning the isotope fractionation taking place during fragmentation in EI-MS is conducted. Both intermolecular and intramolecular chlorine/bromine isotope fractionations present in EI-MS during dehalogenation processes. Generally, intermolecular isotope fractionation is normal for precursor ions but inverse for the corresponding product ions. Molecular ions are only normally affected by one type of intermolecular isotope fractionation during the dehalogenation from parent ions to the product ions. While product ions can be impacted by two type of isotope fractionations, i.e., the inverse intermolecular isotope fractionation from the parent ions to the product ions and the normal intermolecular isotope fractionation from the product ions to their further product ions generated during next-step dehalogenation. For an asymmetric compound possessing position-distinct isotope atoms, if the dehalogenation reacts on only one position, the intermolecular isotope fractionation occurring in this condition is normal for the precursor ion, but with no effect to the product ion in comparison with the total precursor ions. Intramolecular isotope fractionation positively affects the isotope ratios of product ions but has no effect on the precursor ions.

Because intermolecular isotope fractionation always presents during fragmentation processes, the isotopologue distributions of the detected precursor ions never obey binomial distribution, no matter whether the isotopologues distributions of the total precursor ions comply with binomial distribution or not. Therefore, the presently generally used isotope ratio evaluation schemes using pair(s) of neighboring isotopologues are improbable to obtain the isotope ratios exactly equal to those of the complete isotopologues. While using the complete-isotopologue scheme can certainly obtain the isotope ratios of the complete isotopologues. As the isotopologue distribution modes of analytes are anticipated to always be different from those of external isotopic standards, the measured real isotope ratios calculated with the isotopologue-pair scheme probably do not accurately reflect those in reality, despite of application of calibration using external isotopic standards. This deduction has been experimentally validated with the isotopically distinct standards of PCE and TCE from different producers.

## ASSOCIATED CONTENT

The *Supporting Information* is available free of charge on the ACS Publications website at http://pending.

## ACKNOWLEDGEMENTS

We are grateful for Mr. Jianhua Tan, from Guangzhou Quality Supervision and Testing Institute, China for his gift of the chloroethylene standards and valuable suggestions. The authors truly appreciate the assistance from Mr. Ruotai Li (The School of Mathematical Sciences at Peking University, China), Mr. Xianda Sun (Rongxin Power Electronic Co., Ltd., China) and Mr. Ke Zheng (Guangzhou Institute of Geochemistry) in checking the mathematical deduction. This work was partially supported by the National Natural Science Foundation of China (Grant No. 41603092).

**Figure Legends**

**Figure 1.** Illustration of pathways and the probabilities from molecular ion isotopologues to dechlorinatied product ion isotopologues in EI-MS. $x_0$-$x_4$: molar amounts of the molecular ion isotopologues; $I_1$-$I_3$: MS signal intensities of the product ion isotopologues.

**Figure 2**. The structures of the fictitious molecule (M0) containing two chlorine atoms and the four possible isotopomers (MI1, MI2, MI3 and MI4). a: bond a of M0, b: bond b of M0; $a_1$’: bond $a_1$’ of M1, $b_1$’: bond $b_1$’ of M1; $a_2$’: bond $a_2$’ of M2, $b_2$’: bond $b_2$’ of M2; $a_3$’: bond $a_3$’ of M3, $b_3$’: bond $b_3$’ of M3; $a_4$’: bond $a_4$’ of M4, $b_4$’: bond $b_4$’ of M4. The molar amounts of M0, MI1, MI2, MI3 and MI4 are $a_1+a_2+a_3+a_4$, $a_1$, $a_2$, $a_3$ and $a_4$, respectively.

**Figure 3**. The four possible dehalogenation pathways of the MI2 and MI3 (molecular isotopomer 2 and 3) during fragmentation in EI-MS. The molar amounts of MI2 and MI3 reacting in the four dehalogenation pathways are $b_2$, $b_3$, $a_2 – b_2$, and $a_3 – b_3$, respectively. MI2-F1, MI3-F1, MI2-F2 and MI3-F2 are the four possible fragments generated from MI2 and/or MI3 through the four dehalogenation pathways.

**Figure 4**. The four possible dehalogenation pathways of MI1 and MI4 (molecular isotopomer 1 and 4) during fragmentation in EI-MS. The molar amounts of MI1 and MI4 reacting in the four dehalogenation pathways are $b_1$, $b_4$, $a_1 – b_1$, and $a_4 – b_4$, respectively. MI1-F1, MI4-F1, MI1-F2 and MI4-F2 are the four possible fragments generated from MI1 and/or MI4 through the four dehalogenation pathways.

**Figure 5**. Measured chlorine isotope ratios of perchloroethylene (PCE) and trichloroethylene (TCE) calculated with the isotopologue-pair scheme using different pairs of neighboring molecular isotopologues, and those calculated with the complete-isotopologue scheme. IP: isotopologue pair, IR: isotope ratio. The correlations between the measured chlorine isotope ratios (y) and the numbers of heavy isotope atoms (x) are fitted with polynomial functions. Error bars show the standard deviations (1 $\sigma$). The standards PCE and TCE (high performance liquid chromatography grade) were bought from the manufacturer-1. The injection replicates for PCE were five, and six for TCE.

**Figure 6**. Measured chlorine isotope ratios of PCE and TCE from two different manufacturers calculated with the isotopologue-pair scheme using the first pair of neighboring molecular isotopologues, and those calculated with complete-isotopologue scheme. The standards PCE and TCE purchased from the manufacturer-2 were of analytical grade. The standards from different manufactures were analyzed alternately and successively, and the injection replicates were six.

## Figures

**Figure 1.** Illustration of pathways and the probabilities from molecular ion isotopologues to dechlorinatied product ion isotopologues in EI-MS. $x_0$-$x_4$: molar amounts of the molecular ion isotopologues; $I_1$-$I_3$: MS signal intensities of the product ion isotopologues.

**Figure 2**. The structures of the fictitious molecule (M0) containing two chlorine atoms and the four possible isotopomers (MI1, MI2, MI3 and MI4). a: bond a of M0, b: bond b of M0; $a_1$': bond $a_1$' of M1, $b_1$': bond $b_1$' of M1; $a_2$': bond $a_2$' of M2, $b_2$': bond $b_2$' of M2; $a_3$': bond $a_3$' of M3, $b_3$': bond $b_3$' of M3; $a_4$': bond $a_4$' of M4, $b_4$': bond $b_4$' of M4. The molar amounts of M0, MI1, MI2, MI3 and MI4 are $a_1+a_2+a_3+a_4$, $a_1$, $a_2$, $a_3$ and $a_4$, respectively.

**Figure 3**. The four possible dehalogenation pathways of the MI2 and MI3 (molecular isotopomer 2 and 3) during fragmentation in EI-MS. The molar amounts of MI2 and MI3 reacting in the four dehalogenation pathways are $b_2$, $b_3$, $a_2 - b_2$, and $a_3 - b_3$, respectively. MI2-F1, MI3-F1, MI2-F2 and MI3-F2 are the four possible fragments generated from MI2 and/or MI3 through the four dehalogenation pathways.

**Figure 4**. The four possible dehalogenation pathways of MI1 and MI4 (molecular isotopomer 1 and 4) during fragmentation in EI-MS. The molar amounts of MI1 and MI4 reacting in the four dehalogenation pathways are $b_1$, $b_4$, $a_1 - b_1$, and $a_4 - b_4$, respectively. MI1-F1, MI4-F1, MI1-F2 and MI4-F2 are the four possible fragments generated from MI1 and/or MI4 through the four dehalogenation pathways.

**Figure 5**. Measured chlorine isotope ratios of perchloroethylene (PCE) and trichloroethylene (TCE) calculated with the isotopologue-pair scheme using different pairs of neighboring molecular isotopologues, and those calculated with the complete-isotopologue scheme. IP: isotopologue pair, IR: isotope ratio. The correlations between the measured chlorine isotope ratios (y) and the numbers of heavy isotope atoms (x) are fitted with polynomial functions. Error bars show the standard deviations (1 $\sigma$). The standards PCE and TCE (high performance liquid chromatography grade) were bought from the manufacturer-1. The injection replicates for PCE were five, and six for TCE.

**Figure 6**. Measured chlorine isotope ratios of PCE and TCE from two different manufacturers calculated with the isotopologue-pair scheme using the first pair of neighboring molecular isotopologues, and those calculated with complete-isotopologue scheme. The standards PCE and TCE purchased from the manufacturer-2 were of analytical grade. The standards from different manufactures were analyzed alternately and successively, and the injection replicates were six.